# Rapid Development of Efficient Participant-Specific Computational Models of the Wrist

Thor E. Andreassen[1], Taylor P. Trentadue[1,2,3], Andrew R. Thoreson[1]

Kai-Nan An[1], Sanjeev Kakar[4], Kristin D. Zhao[1,3,5]

[1]Assistive and Restorative Technology Laboratory, Mayo Clinic, Rochester, Minnesota, USA
[2]Mayo Clinic Medical Scientist Training Program, Mayo Clinic, Rochester, Minnesota, USA
[3]Mayo Clinic Graduate Program in Biomedical Engineering and Physiology, Mayo Clinic, Rochester, Minnesota, USA
[4]Department of Orthopedic Surgery, Mayo Clinic, Rochester, Minnesota, USA
[5]Department of Physical Medicine and Rehabilitation, Mayo Clinic, Rochester, Minnesota, USA

Corresponding Author:
Kristin D. Zhao
Zhao.kristin@mayo.edu

This is a preprint of the article that has not gone through full peer review.

**Graphical Abstract:**

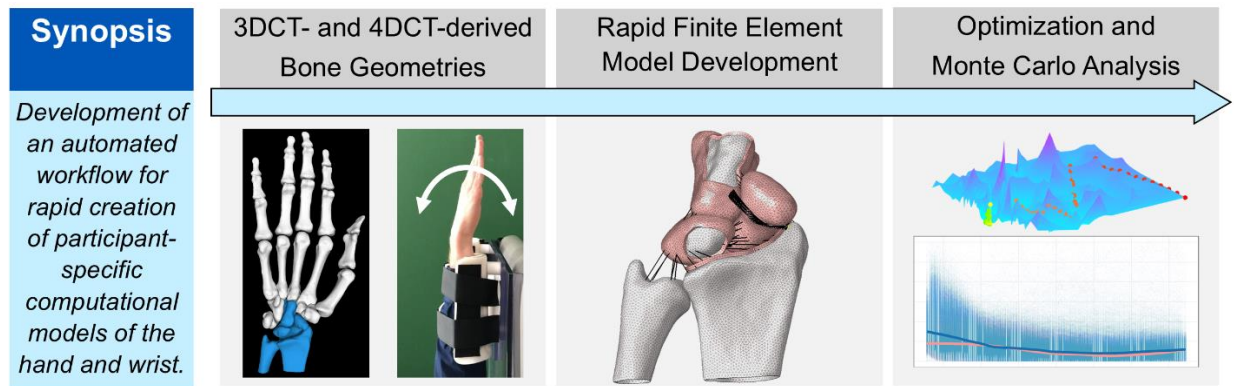

**Abstract:**

While computational modeling may help to develop new treatment options for hand and wrist injuries, at present, few models exist. The time and expertise required to develop and use these models is considerable. Moreover, most do not allow for variation of material properties, instead relying on literature reported averages. We have developed a novel automated workflow combining non-linear morphing techniques with various algorithmic techniques to create participant-specific finite element models. Using this workflow, three participant -specific models were created from our existing four-dimensional computed tomography (4DCT) data. These were then used to perform two analyses to demonstrate the usefulness of the models to investigate clinical questions, namely optimization of ligament properties to participant-specific kinematics, and Monte Carlo (MC) analysis of the impacts of ligament injury on joint contact pressure, as an analogue for joint injury that may lead to osteoarthritis. Participant-specific models can be created in 2 hours and individual simulations performed in 45 seconds. This work lays the groundwork for future patient-specific modeling of the hand and wrist.

# 1 Introduction

Despite comparable numbers of publications between upper and lower limb research (Figure 1A), applications of finite element modeling (FEM) predominate in the lower extremity (Figure 1B). Of the six major appendicular joints, the wrist lags in FEM-based studies [1]. This is problematic, as approximately 5% of individuals experience hand and wrist pain [2], resulting in long occupational absences [3] and high socioeconomic cost [4]. While FEM has informed treatment strategies in other joints [5, 6], the paucity of hand and wrist models limits translational benefit.

For instance, scapholunate (SL) interosseous ligament (SLIL) injuries—among the most common wrist injuries [7, 8]—could be explored using FEM to inform carpal mechanics, progression to SL advanced collapse (SLAC)-pattern osteoarthritis, and clinical decision-making [9]. One study investigated SLIL injuries in the context of lunate morphology, concluding that the type II lunate-hamate facet conferred greater SL joint stability following SLIL injuries [10]. Another compared FEM against experimentally-obtained joint pressure, finding that SLIL injuries created measurable, though not significant, changes in joint contact mechanics that may contribute to osteoarthritis [11].

Still, FEM generalizability may be limited without incorporating individual variation, shown in other joints to improve accuracy [12, 13]. Both aforementioned studies—and most wrist FEM manuscripts—used one or two wrist models with participant-specific geometries, but with literature-derived average material properties without calibration or variation [10, 11, 14-20]. While some have acknowledged this limitation [11, 14, 17], it has not been widely addressed. In the context of American Society of Mechanical Engineers and Food and Drug Administration recommendations, uncertainty quantifications are crucial and recommended for modeling-based studies in medical device evaluation and *in silico* trials [21, 22].

Beyond the necessity of these variations for regulatory use, material variations can provide additional benefits. For example, calibrating material property parameters to experimentally obtained data has improved predictive capabilities [13, 23-25]. Moreover, statistical methods, such as Monte Carlo (MC) analysis give insights into the relationships between parameters and model predictions [26]. However, incorporating these methods in FEM is contingent upon workflows that rapidly develop and execute individualized models. Furthermore, it has been shown that the largest barrier to adopting computational modeling in clinical-translational scenarios is the time required to build and use models [27]. Most workflows require expertise and substantial computational resources, rendering multiple models impractical. While model run-times are not commonly published, single simulations exceeding 242 hours have been reported [15]. Addressing the time-intensive nature of FEM is critical as interest in personalized models expands.

Our overall goal was to develop a workflow capable of rapidly creating participant-specific wrist FEMs with short computation times to allow efficient material property variation. Once developed, we used the workflow to create three participant-specific models based on our existing four-dimensional computed tomography (4DCT) data. Toward material property variation applications, models were used to perform two separate analyses. For the first aim, participant-specific ligament properties were determined via optimization of model-predicted kinematics to experimentally obtained 4DCT kinematics. For the second aim, MC analyses were performed to demonstrate how computationally simulated SL injuries alter radioscaphoid joint contact pressure, while accounting for natural differences in material variation; this was performed in the context of simulating osteoarthritis progression. This work establishes the foundation for participant-specific FEMs in the wrist.

## 2 Methods

FEMs were created in Abaqus Explicit (Dassault Systemes, France) with participant-specific geometries and a range of material properties, boundary conditions, and simulation setups. An overview of the experimental data, details of model development, and analyses performed is outlined (Figure 3). Specific Abaqus implementations are included as Supplementary Material.

### 2.1 Experimental Data

Static and 4DCT wrist images were acquired from asymptomatic participants in an IRB-approved study (NCT03193996) [28, 29]; a subset of three participants is included (Table 1). Static CT volumes spanned the distal radial metadiaphysis through the distal phalanges. Participants were randomized to one of three dynamic tasks: flexion-extension (FE), radial-ulnar deviation (RU), and grip. During 4DCT data collections, bilateral wrists were imaged while performing functional tasks using previously published methods [30]. Participants performed tasks first without resistance (unresisted) and then with moderate resistance (resisted). 4DCT captured 15 timepoints per motion cycle with 66 ms temporal resolution. A motion cycle was defined as moving from one extremum to the other and back [30].

Models were restricted to the radius, scaphoid, lunate, capitate, and ulna. Bones were segmented from static CT in Analyze (Mayo Foundation for Medical Education and Research, Rochester, MN). Segmentations were used to create three-dimensional surfaces (Figure 3), exported as stereolithography mesh files. Coordinate systems were assigned to the radius (RCS) using a modified ISB standard coordinate system [31]. Coordinate systems for other bones were defined relative to the RCS in the static image and translated to the bone centroid. A custom pipeline was used to produce *n* 4×4 homogenous transformation matrices in RCS at each of *n* timepoints per motion arc [32, 33]. The 3D geometries and corresponding kinematics were the basis for the models created herein.

### 2.2 Model Development

The novelty of the current work was the development of a pipeline to create participant-specific FEMs from unique carpal geometries. Model development is divided into four parts: creating geometries, assigning material properties, applying loading and boundary conditions, and model input variation and result extraction.

#### 2.2.1 Geometries

The critical process underlying rapid model development is the automated creation of geometries by leveraging morphing algorithms, followed by processes to predict non-osseous structure geometries.

*2.2.1.1 Bones*

Bone meshes were smoothed and remeshed as rigid triangular element surfaces with edge lengths of approximately 1.5 mm.

*2.2.1.2 Cartilage*

Cartilage geometries were predicted via a two-stage process using MATLAB (R2024a, MathWorks, Natick, MA). In the first stage, cartilage geometries were estimated by morphing a template geometry with known cartilaginous regions from a published model [16] to each participant. In the published dataset, 3D cartilage geometries were projected onto corresponding bones to define cartilage coverage (Figure 4A). An open-source, previously validated morphing algorithm was applied between the target participant and source template geometries [34]. The solution was applied to the projected template cartilage geometry and used to predict coverage (Figure 4B).

In the second stage, these surfaces were used to develop full 3D cartilage elements [35]. A vector field was created from nearest-neighbor distances between covered surfaces (Figure 4C). A generalized regression neural network was used to define a non-linear mapping, which was used to extrude nodes (Figure 4C). These were combined to create 3D triangular wedge elements representing full-thickness cartilage [15] with edge lengths approximately 0.35 mm. In some instances, cartilage surfaces contained overclosures, which were removed using a combination of HyperMesh (Altair Engineering Inc., Troy, MI) and another overclosure algorithm [34]. Mesh quality verification was conducted per recommendations using FEBio [36, 37]; results are in Supplemental Material.

*2.2.1.3    Ligaments*

Ligaments were created in a two-stage process (Figure 5). In the first stage, a template geometry with ligament attachment sites was created for each bone. Attachment sites were identified using previously modeled ligaments [16] as well as manual identification based on anatomical descriptions [38-44] (Table 2). This template was morphed to each participant's geometry, using the processes described for cartilage [34], to predict individualized ligament attachment sites (Figure 5A). Non-linear morphing has low errors in predicting physical landmarks and ligament attachments with minimal user involvement [45, 46].

The number of ligamentous fibers was manually defined (Table 2) to create similar density of fibers across ligament attachment sites. Individual fiber endpoints were determined using a K-means algorithm. Fibers were created by defining correspondence between endpoints on each site using the Kuhn-Munkres algorithm (Figure 5B). The average length of the fibers for each bundle was recorded for use in the material model. Ligaments were exported as 1D non-linear connector elements.

The radioscaphocapitate (RSC), which wraps around the distal scaphoid and has unique properties proximally versus distally [39], was modeled as a 2D fiber-reinforced membrane [47]. It was created using identified ligament attachments on the radius and capitate as well as the approximate scaphoid contact region. A set of interpolated curves connecting these regions was used to create a quadrilateral mesh surface based on the desired element size. The final mesh had quadrilateral membrane elements with 1D non-linear connector elements connected along and across quadrilateral elements, a figure showing this is provided in the Supplemental Material.

### 2.2.2    Material Properties

Rapid model execution is facilitated by sophisticated model material properties that convert time-consuming material non-linearities into tuned linear approximations. These approximations exhibit similar model behavior but with reduced computational overhead.

*2.2.2.1    Bones*

As the focus of this work is recreating bone motion, rather than underlying stress distributions, bone deformation is negligible. Bones were modeled as rigid bodies with their motion restricted to 6DOF motion of a single rigid body node (RBN) at the center of mass.

*2.2.2.2 Cartilage*

Cartilage was modeled as a rigid-rigid material with a tri-linear contact pressure-overclosure relationship defined between surfaces [48, 49] calibrated to the deformation of contacting non-linear hyperelastic materials, with literature-reported values (Mooney-Rivlin material, $C_{10}$ = 4.1, $C_{01}$ = 0.41, $D_1$= 0.0475) [15, 50]. This behaves like a non-linear deformable material but does not permit internal material stress calculations, improving computational speed for contact analysis while permitting contact outcome quantifications [48, 51]. The calibration process is elaborated in Supplemental Material.

*2.2.2.3 Ligaments*

1D ligaments were modeled with a non-linear tension-only force-displacement curve. Material representations are based on previous musculoskeletal modeling [52, 53]. Each ligament is modeled with a literature-derived constant toe-in strain parameter (0.03) [54] and ligament-specific reference length, stiffness, and reference strain. Reference ligament length was defined as the average ligament bundle length. Stiffness and reference strain were design variables, allowing for variation in material properties. Where available, literature-derived mean ±3 standard deviation (SD) ligament stiffness ranges were used [20, 55, 56]. Unavailable SDs were estimated using a coefficient of variation of 0.3.

As reference strain ranges in the wrist are underreported, they were estimated based on work in the knee. As reference strain across all knee ligaments was 0.96 ± 0.13, rarely exceeding 1.30 [13, 57], the parameter ranges were set at 1.00 ± 0.30. Reference strain values exceeding 1.00 represent tension in the base pose of the model, while values under 1.00 represent slack.

All ligament fibers within a bundle were given equivalent material properties. To allow for simulated failure, ligaments were modeled with a parameter equation [51]. Any negative scalar in the parameter equation would result in immediate ligament failure, simulating injury. In the context of this work, ligament injury represents a complete tear of the ligament. For the dorsal SLIL, complete tears have been found in 78% of cases involving the dorsal SLIL injury[58].

### 2.2.3 Loading and Boundary Conditions

Radius and ulna positions were fixed in all DOF, while the lunate and scaphoid were free in all DOF. Capitate motion differed between the two types of analyses performed. In the first Analysis, capitate motion was applied with experimentally obtained kinematics. In the second analysis, capitate motion was applied with idealized joint kinematics. Cartilage and ligament endpoints were rigidly tied to the bone RBNs. A set of nodes aligned to the local coordinate system for each bone were also tied to the RBNs and used to calculate bone kinematics in post-processing.

Models performed two steps: settling and motion. Initial bone positions were defined from static CT. During the settling step, any ligaments injured were given a failure condition and removed from analysis. Initial resting ligament tension, a result of the applied reference strain, was applied to settle positions of the lunate and scaphoid. The capitate was moved from its static CT pose to its appropriate starting position. In the second step, capitate motion was applied; bone kinematics and results were extracted.

*2.2.3.1 Experimentally Obtained Kinematics*

Models recreated experimentally obtained capitate motion from 4DCT. The 4DCT-derived 4×4 transformations were converted to 6DOF capitate kinematics, relative to the radius, and applied using a displacement boundary condition to the capitate rigid body node and the amplitude trajectory for each DOF at every timepoint.

*2.2.3.2    Idealized Simulated Kinematics*

Models applied idealized wrist flexion-extension kinematics to the capitate. A smooth ramped motion was created from maximum flexion to maximum extension. The other two rotational DOF were kept constant at their initial rotation. All rotational DOF are applied using the displacement boundary condition to the rigid body node. To allow the capitate to maintain contact with the proximal row, all translational DOF were left in zero-load control except for the compressive axis, which applied a small load to ensure joint contact.

### 2.2.4    Model Input Variation and Results Extraction

*2.2.4.1    Run Time Material Variation*

A total of 27 ligament design variables - 22 ligament material properties along with 5 parametric scalars for the subset of ligaments designed to simulate injury (Table 2) – were defined for each analysis. Design variables were updated using a previously developed custom automated software interface [13, 25].

*2.2.4.2    Model Output Extraction*

FEM results were extracted using custom software based on Hume et al. [25]. Briefly, a custom set of MATLAB and Python scripts are used to extract model results using the Abaqus API. The MATLAB script then organizes the outputs of interest and calculates transformation matrices for each bone. All captured outputs were combined with the original ligament material design variables for the iteration and appended to a log file of all model runs. This allows future analyses to use all captured values as inputs.

## 2.3    Analyses

Two different analyses demonstrate how models can be applied to clinically relevant questions. In the first analysis, ligament material properties were calibrated to match model predictions of joint motion to experimental data. In the second analysis, ligament material properties were varied using MC analysis to evaluate effects of ligament injury on joint contact pressure while controlling for variation in material properties. All analyses were performed on a standard computer without a dedicated graphics processing unit, using four parallel cores on the central processing unit (Intel Core i5-12600 Processor). The analysis relied on standard hardware on a single machine for broader applicability.

### 2.3.1    Analysis 1 - Participant-Specific Ligament Material Optimization

This analysis demonstrates a method to optimize ligament material properties of each model to best match the observed motion of individuals towards development of patient-specific models. This allows for quantification of individualized loads, joint contact, etc. The analysis includes comparisons of experimentally optimized material properties and model performance of kinematic predictions between the models with optimized material properties and those with average material properties.

*2.3.1.1    Optimization Steps*

Capitate motion was driven with experimentally obtained 4DCT kinematics during the unresisted joint motion, while motion of the scaphoid and lunate were optimized to match their experimentally-obtained motion. Based on previous methods, an error transformation matrix was calculated between the model-observed and the experimentally-obtained bone position for the scaphoid and lunate [59]. This transformation matrix represents the difference in bone position and alignment between the model prediction and experimental measurement. These were converted to 6DOF kinematics. The magnitude of error was calculated for each DOF. The 75$^{th}$ percentile of error across all frames was used as a single scalar estimate of error for each DOF [51]. Errors were normalized by the range of experimental kinematic values for the given DOF. Normalized errors were scaled by corresponding weights based on the relevance of each DOF. The resulting weighted errors were summed for all DOF across both the lunate and scaphoid to estimate overall cost. Optimization algorithms then iteratively adjusted ligament material properties to minimize this cost. Optimization was performed first using a surrogate optimization and then, following convergence, a local optimization based on gradient descent.

*2.3.1.2 Simulations*

Optimization of material properties was performed for all three models to their respective unresisted motion (Table 1). The optimized material properties (reference strain and stiffness) were recorded allowing for investigation of inter-participant differences. Reported values were also normalized to their respective Z-score ranges based on the material property average and standard deviation. A comparison of the model kinematic accuracy against experimental data during the unresisted motion was done, for both the calibrated and uncalibrated models. Then to compare the added robustness of the models with calibrated material properties, a benchmarked activity using the participant's resisted motion, not used for calibration, was evaluated. Root mean squared errors (RMSE) were calculated for each DOF in each model relative to the experimentally obtained motion.

### 2.3.2 Analysis 2 - Ligament Monte Carlo Analysis

This analysis demonstrates a way to use Monte Carlo (MC) analysis to investigate the effects of ligament injuries while accounting for variations in ligament material properties. Simulating injuries to ligamentous stabilizers can inform on sensitivity of radioscaphoid contact pressure to resulting ligament material property variations. Additionally, MC analysis allows investigations into the distribution of ligament forces, something impractical to do with experimental methods alone.

*2.3.2.1 Monte Carlo Steps*

Joint motions were predicted following unique combinations of material variation and ligament injury. The models applied idealized flexion-extension motion to the capitate during each simulation. Each simulation used inputs randomly chosen from each parameter's distribution. For ligaments commonly associated with SL injuries (Table 2), the scalars for injury were randomly sampled, with a 30% chance of failure. Each ligament injury was simulated independently, allowing for all possible combinations to be simulated. Latin hypercube sampling was used to choose configurations. Simulations recorded the combination of injured ligaments and radioscaphoid contact pressure throughout the motion. Upon visual inspection, models with contact pressures exceeding 16.5 MPa yielded unrealistic simulations; 68 (0.27%) were therefore excluded, but simulations were repeated until 12,500 successful simulations were performed.

*2.3.2.2 Simulations*

Models of age- and sex-matched Participants 1 and 2 were included in the MC analysis with a total of 25,000 simulations. Simulations were analyzed using generalized linear mixed effects (GLME) models in R (R Foundation for Statistical Computing, Austria). GLME models predicted radioscaphoid contact pressure as a function of wrist angle and the presence of specific ligament injuries. Wrist flexion-extension angle and ligament injury states were fixed effects. Unique combinations of material properties were treated as random effects. Significance was defined at $\alpha=0.001$. Radioscaphoid contact pressures were compared with previous wrist models [11, 14]. For intact simulations, ligament forces in the volar, membranous, and dorsal SLIL were quantified across the motion. GLME models were used to determine differences between loads in each portion of the SLIL.

# 3 Results

## 3.1 Computational Speed

### 3.1.1 Model Development

Following workflow development, the time required to create a new participant-specific model—excluding bone segmentation time—was under two hours of unsupervised computation.

### 3.1.2 Model Simulation

Each MC iteration took approximately 45 seconds, totaling approximately 13 days of unsupervised computation time for all 25,000 simulations. Material optimization iterations required more time to converge with experimental kinematics, approximately 145 seconds for 275 iterations. The total optimization time was approximately 11 hours for each model.

## 3.2 Analysis 1 - Participant-Specific Ligament Material Optimization

### 3.2.1 Material Properties

Ligament stiffness and reference strain values were generally unique to individuals (Figure 6). When normalized to z-scores, some ligaments were consistently different from reported mean material properties. Volar SLIL values were consistently stiffer than reported values included in Table 2. In addition, for all individuals, the volar SLIL had reference strains greater than 1.0, and the dorsal radiocapitate ligament had reference strains less than 1.0.

### 3.2.2 Participant-Specific Kinematic Errors

Optimization of ligament parameters decreased errors for predicted kinematics for both the original optimized activity (unresisted, Figure 7 and Table 3) and the uncalibrated benchmark activity (resisted, Table 4) but improvement varied across DOF, and were generally better for the scaphoid than the lunate. For the optimized activity, median [$5^{th}$, $95^{th}$ percentile] changes in translational RMSE across all DOF were -0.1 [-0.3, 0.4] and -0.2 [-0.7, 0.5] mm for the scaphoid and lunate, respectively. Changes in rotational RMSE across all DOF were -1.2 [-4.0, 1.3] and -0.4 [-5.8, 1.7] for the scaphoid and lunate, respectively. For the uncalibrated benchmarked activity, median changes in translational RMSE were 0.0 [-0.6, 0.5] and -0.1 [-0.7, 0.8] mm for the scaphoid and lunate. Changes in rotational RMSE were -0.8 [-2.3, 0.9] and -0.1 [-3.3, 2.3] for the scaphoid and lunate. Negative values represent superior accuracy of the optimized over uncalibrated models.

## 3.3 Analysis 2 - Ligament Property Monte Carlo Analysis

### 3.3.1 Radioscaphoid Contact Pressure

Adjusting for ligament variability, SL ligament injuries and wrist extension increased radioscaphoid contact pressure (Figure 8A, Table 5). The RSC ligament was associated with a statistically significant 46.5% increase in radioscaphoid contact pressure relative to the average joint pressure independent of material properties, model, and integrity of other ligaments (Figure 8B, Table 5). Injury to the dorsal SLIL was not associated with a significant change in radioscaphoid contact pressure (Table 5). Radioscaphoid contact pressures were compared to published work (Figure 9).

### 3.3.2 Ligament Forces

Ligament forces in the intact dorsal and volar SLIL increased with increasing wrist extension angle, while force in the membranous SLIL remained relatively constant throughout motion (Figure 10). When accounting for material property variation, the average ligament force in the volar SLIL was significantly less than the dorsal SLIL by 1.6 N (Table 6). Dorsal and volar SLIL loads were 25.7 N and 24.1 N greater, respectively, than the proximal SLIL.

## 4   Discussion

Computational modeling is poised to enable new insights into treatment for hand and wrist injuries, as it has for other joints; however, this has been stifled by a lack of models. Hand and wrist models lag all other appendicular joints for FEM-based studies. For those models that do exist, they do not allow for individual material variation as their development and use remains inefficient. Improving this time limitation is crucial to increasing the personalization of these models and to increasing adoption of FEM for hand and wrist research.

To this end, we have developed workflows to rapidly generate participant-specific wrist models that permit material property variation. The workflow was nearly entirely automated, relying on morphing techniques and algorithmic predictions that can be executed in under two hours. The resulting models can be run on a standard computer in 45 seconds. With this workflow, we developed three participant-specific models and used them in two analyses to demonstrate validity and potential applications.

In the first analysis, ligament material properties for participant-specific models were optimized to match an individual's 4DCT-derived joint motion, which captured a task performed at speeds physiologically relevant to daily activities. Ligament properties varied between individuals following optimization (Figure 6) yet there were some trends. While the sample size is small, the consistent patterns for the volar SLIL and the dorsal radiocapitate (DRC) ligament across all models suggest that some ligaments may be consistently under residual stress (volar SLIL) in a neutral wrist position, while others are slack (DRC). This demonstrates how participant-specific models could be used to supplement traditional material testing by offering insight into *in situ* material performance often impractical to measure.

Additionally, the first analysis showed that optimization improved accuracy of scaphoid and lunate kinematic predictions in both the optimized (unresisted) and the benchmark (resisted) trials across the majority of DOF (Table 3 and Table 4), with greater improvements in optimized (unresisted) trials. In optimized trials, RMSE decreased in 66.7% of DOF across all participants relative to the models with average material properties; however, this decreased to 61.1% for the non-optimized benchmark trials (Table 3 and Table 4). Still, the majority of DOF were improved for the benchmark case, highlighting that optimization improves model robustness and improves the ability of models to make general predictions. In addition, this work demonstrates that the models created can reproduce complicated joint dynamics (Figure 7).

In the second analysis, models of two age- and sex- matched participants were used in a MC analysis to determine the effects of ligament injury on radioscaphoid joint contact pressure. A noticeable increase in radioscaphoid contact pressure was observed with some SL ligament injuries (Figure 7A), but the effect was not consistent between ligaments (Table 5 and Figure 8B). Injury to the RSC caused a significant increase in joint contact pressure even with all other ligaments remaining intact. In contrast, injury to the dorsal SLIL, considered the most crucial SL joint stabilizer [60], caused no significant change in radioscaphoid joint contact pressure. The results of this work emphasize that certain metrics may be more directly associated with certain soft tissue structures. Further, this suggests that average changes to contact behavior may not be the most appropriate biomarker for wrist osteoarthritis progression. A strength of FEM is the ability to investigate changes in multiple metrics that would be impractical with other methods. Still, results should be interpreted in the context of experimental conditions and clinical experience.

The large number of simulations generated results that were statistically significant but not necessarily by an amount that would be physically meaningful; for instance, ligament load in the volar versus dorsal SLIL differed by only 1.6 N (Table 6). While this difference is by itself not likely clinically relevant, it may explain why the volar SLIL is frequently injured before the dorsal SLIL: while forces are nearly identical, failure strength of the volar SLIL is considerably lower [61]. In the intact wrist, while ligament forces continued to increase in wrist extension (Figure 10), median radioscaphoid contact pressure increased until 30° extension and then stabilized (Figure 8A). In contrast, for models with some form of ligament injury, median contact pressure continued to increase with greater wrist extension (Figure 8A). Soft tissue injuries may be more severe at extrema [30], with implications for osteoarthritis development.

These findings generally agree with previous work. As seen in our calibration testing, increased accuracy with use of participant-specific material property values has been demonstrated previously [13, 23, 62]. Still, differences in kinematic accuracies across the models (Table 3 and Table 4) suggest that calibration could be improved. Certain joint motions better inform specific ligament behaviors and, depending on the intended model use, are important to achieve adequate accuracy. In the second analysis, intact radioscaphoid contact pressures overlap those reported in Kim et al. [11] but not Mena et al. [14]; however, average values were less than reported elsewhere (Figure 9). Radioscaphoid pressures were also similar to Kim et al. for the intact, partial, and fully injured SLIL but ranges observed herein were larger. Additionally, ligament forces observed in the MC analysis agree with prior cadaveric testing, which showed that maximum SLIL force was observed at maximum extension, with an average total SLIL force of 20.0 ± 3.3 N at 30° extension [63]. Our models similarly predicted the highest loads with extension, with an average total SLIL force of 26.0 N at 30° extension (Figure 10).

This work is not free of limitations. First, only three carpal bones were included. Other models include all carpal bones and often include more ligaments. Some differences in kinematics likely resulted from the absence of certain bones and ligaments. In the first analysis, the improvements in the scaphoid were superior to the lunate, and kinematics for some parts of the time domain could not be recreated by the model (Figure 7). These differences may be improved with additional structures. In the second analysis, including additional ligaments could change patterns and interpretation. We chose to use a limited set of bones, based on availability of their kinematics from previously tracked 4DCT data [30]. Previous studies have used the same subset of bones to develop a model of radius contact pressures and ligament stress in the SLIL [16]. Crucially, the workflow we have developed is not unique to these bones and could be more broadly applied.

Another limitation is the narrow scope of the validation of the work herein. In the first analysis, the non-optimized trials were used as a benchmark to validate the accuracy of the models during unseen kinematics, testing the extent to which models can be used for general prediction. While this mimics validation protocols for other joints [24], we were unable to directly validate the contact pressures and ligament loads, as there is no practical method to measure these *in vivo*. Still, in the second analysis, we validated that the model predictions for both radioscaphoid contact pressure and ligament loads are consistent with existing work.

Another limitation is the comparison of radioscaphoid contact pressures to those reported in the literature. Previous studies have reported maximum contact pressure at a single node while this work reported average contact pressure [11, 14]. We chose to use the average pressure, as it is less affected by surface reconstruction error, mesh discretization, and choice of contact model than maximum pressure, particularly for rigid-rigid models with pressure overclosure [64, 65].

Last, this study has a relatively small sample size of three models (two for the second analysis), insufficient to make broad claims about the overall behavior of ligament material properties and their effects on contact pressure. However, the goal of this work was first to develop a workflow that would demonstrate superior efficiency for model development and use over existing modeling work in the wrist. The investigations performed into evaluating wrist behavior were meant to highlight some potential uses of these improved models, as opposed to comprehensively evaluating wrist behavior. Future work may expand this to larger samples to rigorously investigate these questions.

In conclusion, we have developed a novel automated workflow to rapidly create participant-specific wrist models that have efficient computational speeds. This workflow was used to develop three participant-specific models, that were then applied in two analyses to demonstrate the relevance of rapid models for clinical purposes. The first analysis showed a method to optimize ligament material properties to match 4DCT-derived kinematics demonstrating a path towards patient-specific models, with potential uses in developing digital twins, pre-surgical planning tools, and in *in silico* medical device trials. The second analysis used models to perform MC simulations, showing how models may additionally be used to investigate the sensitivity of clinical metrics to individual parameters and help explain clinical phenomena. In this case, demonstrating that SL ligament injuries impact radioscaphoid contact pressure and that volar SLILs may be more prone to injury versus dorsal SLILs. These improve our understanding of short- and long-term effects of injury and may guide surgical treatment. Overall, we present a pipeline towards efficient participant-specific wrist modeling.

## 5 Contributor Role Taxonomy (CRediT) Statement

**TEA:** Conceptualization, Formal analysis, Investigation, Methodology, Software, Visualization, Writing - Original Draft Preparation **TPT:** Investigation, Validation, Data Curation, Writing – Original Draft Preparation **ART:** Conceptualization, Validation, Data Curation, Resources, Project Administration **KA:** Conceptualization, Validation, Supervision **SK:** Validation, Supervision **KDZ:** Conceptualization, Validation, Funding Acquisition, Resources, Supervision. All authors performed Writing -Review and Editing.

## 6 Declaration of Competing Interest

Sanjeev Kakar, M.D., received royalties or licenses from Arthrex, which was not related to this work. Other authors declare no known competing financial interests that may influence the work reported in this manuscript.

## 7 Funding

The authors would like to thank the National Institutes Health for providing financial support through the following grants: T32 AR056950 (TEA and TPT), F31 AR082227 (TPT), R01 AR071338 (KDZ), T32 GM065841 (TPT), and T32 GM145408 (TPT). This work was also supported through the Early-Stage Investigator Research Award from the Mayo Clinic Office of Core Shared Services.

## 8 Acknowledgements

The authors thank Altair Engineering Inc. for generously providing HyperWorks software to enable this work. The authors thank David Holmes III, Ph.D., and Jon Camp at the Biomedical Imaging Research Core Facility at the Mayo Clinic for their help with this work. The authors thank Dr. Landon Hamilton at the Medical Center of the Rockies for his help with the statistical analysis in this work.

# 9 References

[1] Mena A, Wollstein R, Baus J, Yang J. Finite Element Modeling of the Human Wrist: A Review. J Wrist Surg. 2023;12:478-87.
[2] Walker-Bone K, Palmer KT, Reading I, Coggon D, Cooper C. Prevalence and impact of musculoskeletal disorders of the upper limb in the general population. Arthritis Rheum. 2004;51:642-51.
[3] Barr AE, Barbe MF, Clark BD. Work-related musculoskeletal disorders of the hand and wrist: epidemiology, pathophysiology, and sensorimotor changes. J Orthop Sports Phys Ther. 2004;34:610-27.
[4] de Putter CE, Selles RW, Polinder S, Panneman MJ, Hovius SE, van Beeck EF. Economic impact of hand and wrist injuries: health-care costs and productivity costs in a population-based study. J Bone Joint Surg Am. 2012;94:e56.
[5] Heller MO. Finite element analysis in orthopedic biomechanics.  Human Orthopaedic Biomechanics2022. p. 637-58.
[6] Favre P, Maquer G, Henderson A, Hertig D, Ciric D, Bischoff JE. In Silico Clinical Trials in the Orthopedic Device Industry: From Fantasy to Reality? Ann Biomed Eng. 2021;49:3213-26.
[7] Ward PJ, Fowler JR. Scapholunate Ligament Tears: Acute Reconstructive Options. Orthop Clin North Am. 2015;46:551-9.
[8] Wang WL, Abboudi J, Gallant G, Jones C, Kirkpatrick W, Kwok M, et al. Radiographic Incidence and Functional Outcomes of Distal Radius Fractures Undergoing Volar Plate Fixation With Concomitant Scapholunate Widening: A Prospective Analysis. Hand (N Y). 2022;17:326-30.
[9] Watson HK, Ballet FL. The SLAC wrist: scapholunate advanced collapse pattern of degenerative arthritis. J Hand Surg Am. 1984;9:358-65.
[10] Leonardo-Diaz R, Alonso-Rasgado T, Jimenez-Cruz D, Bailey CG, Talwalkar S. Performance evaluation of surgical techniques for treatment of scapholunate instability in a type II wrist. Int J Numer Method Biomed Eng. 2020;36:e3278.
[11] Kim S, Nyaaba W, Mzeihem M, Mejia A, Gonzalez M, Amirouche F. Wrist Kinetics Post-Scapholunate Dissociation: Experimental and Computational Analysis of Scapholunate Interosseous Ligament Injury. Journal of Hand Surgery Global Online. 2024.
[12] Gardiner JC. Computational Modeling of Ligament Mechanics. 2002. p. 1-179.
[13] Andreassen TE, Laz PJ, Erdemir A, Besier TF, Halloran JP, Imhauser CW, et al. Deciphering the "Art" in Modeling and Simulation of the Knee Joint: Assessing Model Calibration Workflows and Outcomes. Journal of Biomechanical Engineering2023. p. 1-13.
[14] Mena A, Wollstein R, Yang J. Development of a Finite Element Model of the Human Wrist Joint with Radial and Ulnar Axial Force Distribution and Radiocarpal Contact Validation. J Biomech Eng. 2025:1-39.
[15] Gislason MK, Stansfield B, Nash DH. Finite element model creation and stability considerations of complex biological articulation: The human wrist joint. Med Eng Phys. 2010;32:523-31.
[16] Marques R, Melchor J, Sanchez-Montesinos I, Roda O, Rus G, Hernandez-Cortes P. Biomechanical Finite Element Method Model of the Proximal Carpal Row and Experimental Validation. Front Physiol. 2021;12:749372.
[17] Alonso Rasgado T, Zhang Q, Jimenez Cruz D, Bailey C, Pinder E, Mandaleson A, et al. Analysis of tenodesis techniques for treatment of scapholunate instability using the finite element method. Int J Numer Method Biomed Eng. 2017;33.
[18] Guo X, Fan Y, Li ZM. Effects of dividing the transverse carpal ligament on the mechanical behavior of the carpal bones under axial compressive load: a finite element study. Med Eng Phys. 2009;31:188-94.

[19] Varga P, Schefzig P, Unger E, Mayr W, Zysset PK, Erhart J. Finite element based estimation of contact areas and pressures of the human scaphoid in various functional positions of the hand. J Biomech. 2013;46:984-90.

[20] Bajuri MN, Abdul Kadir MR, Murali MR, Kamarul T. Biomechanical analysis of the wrist arthroplasty in rheumatoid arthritis: a finite element analysis. Med Biol Eng Comput. 2013;51:175-86.

[21] ASME. Assessing Credibility of Computational Modeling through Verification and Validation: Application to Medical Devices. V&V40: ASME; 2018. p. 1-48.

[22] FDA. Assessing the Credibility of Computational Modeling and Simulation in Medical Device Submissions Guidance for Industry and Food and Drug Administration Staff In: Health CfDaR, editor.2023. p. 1-42.

[23] Anantha Krishnan A, Myers CA, Scinto M, Marshall BN, Clary CW. Specimen-specific finite element representations of implanted hip capsules. Comput Methods Biomech Biomed Engin. 2024;27:751-64.

[24] Erdemir A, Besier TF, Halloran JP, Imhauser CW, Laz PJ, Morrison TM, et al. Deciphering the "Art" in Modeling and Simulation of the Knee Joint: Overall Strategy. Journal of Biomechanical Engineering2019. p. 1-10.

[25] Hume DR, Rullkoetter PJ, Shelburne KB. ReadySim: A computational framework for building explicit finite element musculoskeletal simulations directly from motion laboratory data. International Journal for Numerical Methods in Biomedical Engineering2020. p. 1-11.

[26] Baldwin MA, Laz PJ, Stowe JQ, Rullkoetter PJ. Efficient probabilistic representation of tibiofemoral soft tissue constraint. Computer Methods in Biomechanics and Biomedical Engineering2009. p. 651-9.

[27] Lesage R, Van Oudheusden M, Schievano S, Van Hoyweghen I, Geris L, Capelli C. Mapping the use of computational modelling and simulation in clinics: A survey. Frontiers in Medical Technology2023. p. 1-10.

[28] Trentadue TP, Thoreson A, Lopez C, Breighner RE, Leng S, Holmes DR, 3rd, et al. Morphology of the scaphotrapeziotrapezoid joint: A multi-domain statistical shape modeling approach. J Orthop Res. 2024;42:2562-74.

[29] Trentadue TP, Thoreson A, Lopez C, Breighner RE, Leng S, Kakar S, et al. Sex differences in photon-counting detector computed tomography-derived scaphotrapeziotrapezoid joint morphometrics. Skeletal Radiol. 2025.

[30] Trentadue TP, Thoreson AR, Lopez C, Breighner RE, An KN, Holmes DR, 3rd, et al. Detection of scapholunate interosseous ligament injury using dynamic computed tomography-derived arthrokinematics: A prospective clinical trial. Med Eng Phys. 2024;128:104172.

[31] Coburn JC, Upal MA, Crisco JJ. Coordinate systems for the carpal bones of the wrist. J Biomech. 2007;40:203-9.

[32] Zhao K, Breighner R, Holmes D, 3rd, Leng S, McCollough C, An KN. A technique for quantifying wrist motion using four-dimensional computed tomography: approach and validation. J Biomech Eng. 2015;137:0745011-5.

[33] Trentadue TP, Lopez C, Breighner RE, Fautsch K, Leng S, Holmes Iii DR, et al. Evaluation of Scapholunate Injury and Repair with Dynamic (4D) CT: A Preliminary Report of Two Cases. J Wrist Surg. 2023;12:248-60.

[34] Andreassen TE, Hume DR, Hamilton LD, Higinbotham SE, Shelburne KB. Automated 2D and 3D finite element overclosure adjustment and mesh morphing using generalized regression neural networks. Medical Engineering & Physics. 2024;126.

[35] Marai GE, Crisco JJ, Laidlaw DH. A kinematics-based method for generating cartilage maps and deformations in the multi-articulating wrist joint from CT images. Conf Proc IEEE Eng Med Biol Soc. 2006;2006:2079-82.

[36] Maas SA, Ellis BJ, Ateshian GA, Weiss JA. FEBio: Finite elements for biomechanics.  Journal of Biomechanical Engineering2012. p. 1-10.
[37] Anderson AE, Ellis BJ, Weiss JA. Verification, validation and sensitivity studies in computational biomechanics.  Computer Methods in Biomechanics and Biomedical Engineering2007. p. 171-84.
[38] Apergis E. Wrist Anatomy.  Fracture-Dislocations of the Wrist2013. p. 7-41.
[39] Berger RA. The anatomy of the ligaments of the wrist and distal radioulnar joints. Clin Orthop Relat Res. 2001:32-40.
[40] Mizuseki T, Ikuta Y. The dorsal carpal ligaments: their anatomy and function. J Hand Surg Br. 1989;14:91-8.
[41] Moritomo H. Anatomy and clinical relevance of the ulnocarpal ligament. J Wrist Surg. 2013;2:186-9.
[42] Nagao S, Patterson RM, Buford WL, Jr., Andersen CR, Shah MA, Viegas SF. Three-dimensional description of ligamentous attachments around the lunate. J Hand Surg Am. 2005;30:685-92.
[43] Ringler MD. MRI of wrist ligaments. J Hand Surg Am. 2013;38:2034-46.
[44] Zhang J, Yao X, Song Y, Yin P. Establishment and preliminary evaluation of CT-based classification for distal radius fracture. Sci Rep. 2024;14:9673.
[45] Malbouby V, Gibbons KD, Bursa N, Ivy AK, Fitzpatrick CK. Efficient development of subject-specific finite element knee models: Automated identification of soft-tissue attachments. J Biomech. 2025;178:112441.
[46] Pillet H, Bergamini E, Rochcongar G, Camomilla V, Thoreux P, Rouch P, et al. Femur, tibia and fibula bone templates to estimate subject-specific knee ligament attachment site locations.  Journal of Biomechanics: Elsevier; 2016. p. 3523-8.
[47] Baldwin MA, Clary CW, Fitzpatrick CK, Deacy JS, Maletsky LP, Rullkoetter PJ. Dynamic finite element knee simulation for evaluation of knee replacement mechanics. J Biomech. 2012;45:474-83.
[48] Huff DN, Myers CA, Rullkoetter PJ. Impact of alignment and kinematic variation on resistive moment and dislocation propensity for THA with lipped and neutral liners. Biomech Model Mechanobiol. 2020;19:1297-307.
[49] Halloran JP, Petrella AJ, Rullkoetter PJ. Explicit finite element modeling of total knee replacement mechanics.  Journal of Biomechanics2005. p. 323-31.
[50] Li Z, Kim JE, Davidson JS, Etheridge BS, Alonso JE, Eberhardt AW. Biomechanical response of the pubic symphysis in lateral pelvic impacts: a finite element study. J Biomech. 2007;40:2758-66.
[51] Andreassen TE, Hume DR, Hamilton LD, Hegg SL, Higinbotham SE, Shelburne KB. Validation of Subject-Specific Knee Models from In Vivo Measurements. 2024:1-34.
[52] Blankevoort L, Kuiper JH, Huiskes R, Grootenboer HJ. Articular contact in a three-dimensional model of the knee. J Biomech. 1991;24:1019-31.
[53] Yu CH, Walker PS, Dewar ME. The effect of design variables of condylar total knees on the joint forces in step climbing based on a computer model.  Journal of Biomechanics2001. p. 1011-21.
[54] Butler DL, Kay MD, Stouffer DC. Comparison of material properties in fascicle-bone units from human patellar tendon and knee ligaments. J Biomech. 1986;19:425-32.
[55] Chang N, Suh N. Anatomy and Biomechanics of Scapholunate Ligament.  Arthroscopy and Endoscopy of the Elbow, Wrist and Hand2022. p. 633-40.
[56] Nikolopoulos FV, Apergis EP, Poulilios AD, Papagelopoulos PJ, Zoubos AV, Kefalas VA. Biomechanical properties of the scapholunate ligament and the importance of its portions in the capitate intrusion injury. Clin Biomech (Bristol, Avon). 2011;26:819-23.
[57] Zhang Q, Adam NC, Hosseini Nasab SH, Taylor WR, Smith CR. Techniques for In Vivo Measurement of Ligament and Tendon Strain: A Review.  Annals of Biomedical Engineering2021. p. 7-28.
[58] Andersson JK, Garcia-Elias M. Dorsal scapholunate ligament injury: a classification of clinical forms. J Hand Surg Eur Vol. 2013;38:165-9.

[59] Hamilton LD, Andreassen TE, Myers C, Shelburne KB, Clary C, Rullkoetter PJ. Supine leg press as an alternative to standing lunge in high-speed stereo radiography.  Journal of Biomechanics: Elsevier Ltd; 2022. p. 111118.
[60] Berger RA. The gross and histologic anatomy of the scapholunate interosseous ligament. J Hand Surg Am. 1996;21:170-8.
[61] Berger RA, Imeada T, Berglund L, An KN. Constraint and material properties of the subregions of the scapholunate interosseous ligament. J Hand Surg Am. 1999;24:953-62.
[62] Razu SS, Jahandar H, Zhu A, Berube EE, Manzi JE, Pearle AD, et al. Bayesian Calibration of Computational Knee Models to Estimate Subject-Specific Ligament Properties, Tibiofemoral Kinematics, and Anterior Cruciate Ligament Force With Uncertainty Quantification.  Journal of biomechanical engineering2023. p. 1-11.
[63] Dimitris C, Werner FW, Joyce DA, Harley BJ. Force in the Scapholunate Interosseous Ligament During Active Wrist Motion. J Hand Surg Am. 2015;40:1525-33.
[64] Systemes D. ABAQUS Analysis User's Manual. 2020.
[65] Wachman K, Saputra E, Ismail R, Jamari J, Bayuseno AP. Analysis of the Contact Area of Smooth and Rough Surfaces in Contact with Sphere Indenter Using Finite Element Method. MATEC Web of Conferences. 2016;58.

## Tables

Table 1: Participants recruited to a study of normative wrist anatomy and biomechanics were imaged using three- and four-dimensional computed tomography (4DCT) [28, 29]. Three participants were used to generate participant-specific finite element models. The dominant, right hands were modeled for all participants. Type I facets do not have an additional articulation between the lunate and hamate, while Type II facets have an articulation between the lunate and the hamate.

| Participant Number | Sex | Age (years) | Lunate-Hamate Facet Type | 4DCT Motion |
|---|---|---|---|---|
| 1 | Female | 39 | Type I | Radial/ulnar deviation |
| 2 | Male | 38 | Type I | Grip |
| 3 | Female | 29 | Type II | Flexion/extension |

Table 2: Eleven ligaments were included in the model. Information about the ligament design and material properties are detailed. Parameter values are reported as mean ± standard deviation.

| Ligament Name | Ligament Abbreviation | 1D or 2D | Number of Fibers | Stiffness (N/mm) | Reference Strain | Allowed Failure during MC | Material Properties Sources | Ligament Descriptions Sources |
|---|---|---|---|---|---|---|---|---|
| volar scapholunate interosseous ligament | SLIL_v | 1D | 7 | 33±10 | 1.0±0.1 | Yes | Nikolopoulos et al. 2011 | Berger et al. 2001 |
| proximal (membranous) scapholunate interosseous ligament | SLIL_p | 1D | 7 | 25±7.5 | 1.0±0.1 | Yes | Nikolopoulos et al. 2011 | Berger et al. 2001 |
| dorsal scapholuante interosseous ligament | SLIL_d | 1D | 7 | 79±24 | 1.0±0.1 | Yes | MetaAnalysis of values in Chang et al. 2022 | Berger et al. 2001 |
| radioscaphocapitate ligament | RSC | 2D | N/A | 10±3 | 1.0±0.1 | Yes | | Berger et al. 2001 |
| long radiolunate ligament | LRL | 1D | 4 | 75±23 | 1.0±0.1 | Yes | Bajuri et al. 2013 | Berger et al. 2001, Nagao et al. 2005, Zhang et al. 2024 |
| short radiolunate ligament | SRL | 1D | 4 | 75±23 | 1.0±0.1 | No | Bajuri et al. 2013 | Nagao et al. 2005 |
| dorsal radiocarpal ligament | DRC | 1D | 3 | 75±23 | 1.0±0.1 | No | | Nagao et al. 2005, Mizuseki et al. 1989 |
| radial collateral ligament | RCL | 1D | 3 | 10±3 | 1.0±0.1 | No | Bajuri et al. 2013 | Ringler et al. 2013 |
| scaphocapitate ligament | SCL | 1D | 7 | 40±12 | 1.0±0.1 | No | Bajuri et al. 2013 | Apergis et al. 2013 |
| ulnocapitate ligament | UCL | 1D | 3 | 75±23 | 1.0±0.1 | No | | Moritomo et al. 2013 |
| ulnolunate ligament | UL | 1D | 3 | 40±12 | 1.0±0.1 | No | Bajuri et al. 2013 | Nagao et al. 2005 |

Table 3: Root mean squared error (RMSE) of participant-specific model kinematic predictions with "population" average material properties and optimized material properties against experimentally-obtained kinematics during the optimized unresisted joint motion. Dis=distal, Dor=dorsal, Pro=proximal, Rad=radial, Uln=ulnar, Vol=volar.

| Participant | Bone | Material | Translation RMSE (mm) | | | Rotation RMSE (°) | | |
|---|---|---|---|---|---|---|---|---|
| | | | Vol/Dor | Pro/Dis | Rad/Uln | Flex/Ext | UlnD/RadD | Pro/Sup |
| 1 | Scaphoid | Population | 0.31 | 0.52 | 0.23 | 3.84 | 6.13 | 2.62 |
| | | Optimized | 0.22 | 0.41 | 0.27 | 2.21 | 2.13 | 1.45 |
| | Lunate | Population | 0.93 | 0.46 | 0.69 | 5.11 | 3.25 | 2.96 |
| | | Optimized | 1.33 | 0.55 | 0.69 | 5.70 | 3.68 | 4.26 |
| 2 | Scaphoid | Population | 0.45 | 0.43 | 0.31 | 4.54 | 1.03 | 2.32 |
| | | Optimized | 0.13 | 0.09 | 0.02 | 0.99 | 0.23 | 0.32 |
| | Lunate | Population | 0.86 | 0.02 | 0.22 | 5.21 | 5.14 | 2.66 |
| | | Optimized | 0.91 | 0.02 | 0.05 | 5.13 | 5.29 | 1.39 |
| 3 | Scaphoid | Population | 0.43 | 1.66 | 1.21 | 9.55 | 5.81 | 3.76 |
| | | Optimized | 0.97 | 0.93 | 0.78 | 5.75 | 3.09 | 3.52 |
| | Lunate | Population | 1.31 | 0.21 | 0.95 | 10.93 | 1.19 | 3.07 |
| | | Optimized | 0.65 | 0.10 | 0.31 | 5.14 | 2.88 | 2.72 |

Table 4: Root mean squared error (RMSE) of participant-specific model kinematic predictions with "population" average material properties and optimized material properties against experimentally obtained kinematics during the benchmark resisted joint motion. Dis=distal, Dor=dorsal, Pro=proximal, Rad=radial, Uln=ulnar, Vol=volar.

| Participant | Bone | Material | Translation RMSE (mm) | | | Rotation RMSE (°) | | |
|---|---|---|---|---|---|---|---|---|
| | | | Vol/Dor | Pro/Dis | Rad/Uln | Flex/Ext | UlnD/RadD | Pro/Sup |
| 1 | Scaphoid | Population | 0.67 | 0.67 | 0.67 | 4.47 | 4.68 | 3.74 |
| | | Optimized | 0.49 | 0.67 | 0.65 | 3.68 | 2.75 | 3.10 |
| | Lunate | Population | 1.14 | 0.51 | 0.76 | 6.49 | 3.31 | 3.07 |
| | | Optimized | 1.68 | 0.60 | 1.08 | 6.67 | 4.23 | 3.38 |
| 2 | Scaphoid | Population | 1.02 | 1.14 | 0.67 | 9.40 | 3.38 | 6.22 |
| | | Optimized | 0.76 | 0.56 | 0.33 | 7.11 | 1.57 | 3.96 |
| | Lunate | Population | 1.15 | 0.02 | 0.69 | 8.48 | 6.57 | 6.34 |
| | | Optimized | 1.47 | 0.11 | 0.24 | 10.73 | 6.45 | 5.74 |
| 3 | Scaphoid | Population | 0.23 | 1.29 | 0.91 | 4.78 | 7.13 | 2.39 |
| | | Optimized | 1.03 | 0.83 | 0.53 | 3.40 | 3.80 | 2.51 |
| | Lunate | Population | 0.58 | 0.11 | 1.15 | 8.30 | 2.09 | 2.97 |
| | | Optimized | 0.44 | 0.08 | 0.44 | 5.60 | 2.26 | 3.64 |

Table 5: Generalized linear mixed effects (GLME) models were used to predict radioscaphoid contact pressure from ligament injury states in Participants 1 and 2. Values reported are coefficient estimates in true value (MPa) and normalized to average radioscaphoid contact pressure. Positive values represent increased pressure. Probabilistic metrics reported include standard error, t-statistic, and p value. * denotes statistical significance at α=0.001.

|  | Parameter | Estimate - True Value (MPa) | Estimate - Normalized to Average | SE | tStat | pValue |
|---|---|---|---|---|---|---|
| FE Angle Spline | Intercept | 1.294 | 198.0% | 0.006 | 232.01 | <0.0001* |
|  | Capitate Angle DOF 1 | -1.822 | -278.8% | 0.009 | -213.35 | <0.0001* |
|  | Capitate Angle DOF 2 | -0.298 | -45.5% | 0.006 | -52.19 | <0.0001* |
| Ligament Injury | Volar SLIL | 0.003 | 0.4% | 0.005 | 4.99 | <0.0001* |
|  | Proximal SLIL | 0.007 | 1.0% | 0.005 | 13.33 | <0.0001* |
|  | Dorsal SLIL | 0.001 | 0.2% | 0.005 | 2.09 | 0.0363 |
|  | RSC | 0.304 | 46.5% | 0.005 | 59.46 | <0.0001* |
|  | LRL | -0.005 | -0.7% | 0.005 | -8.99 | <0.0001* |

Table 6: Generalized linear mixed effects (GLME) models for SLIL region ligament forces from Monte Carlo simulations with all ligaments intact. The base case is dorsal SLIL; results represent differences from the base case. Positive values represent an increase in ligament force. Probabilistic metrics are standard error, t-statistic, and p value. * denotes statistical significance at α=0.001. The random effect is included since within-model results are not independent measures.

| | Parameter | Estimate (N) | SE | tStat | pValue |
|---|---|---|---|---|---|
| FE Angle Spline | Intercept | 27.840 | 0.175 | 159.45 | <0.0001* |
| | Capitate Angle DOF 1 | -42.580 | 0.167 | -254.37 | <0.0001* |
| | Capitate Angle DOF 2 | -6.613 | 0.111 | -59.64 | <0.0001* |
| Ligament | Proximal SLIL | -25.650 | 0.020 | -1303.91 | <0.0001* |
| | Volar SLIL | -1.576 | 0.008 | -202.76 | <0.0001* |
| | Random Scalar | 0.009 | 0.011 | 0.86 | 0.3910 |
| | Capitate Angle DOF 1 Proximal Interaction | 44.800 | 0.043 | 1039.25 | <0.0001* |
| | Capitate Angle DOF 2 Proximal Interaction | 4.585 | 0.024 | 194.82 | <0.0001* |

Figures

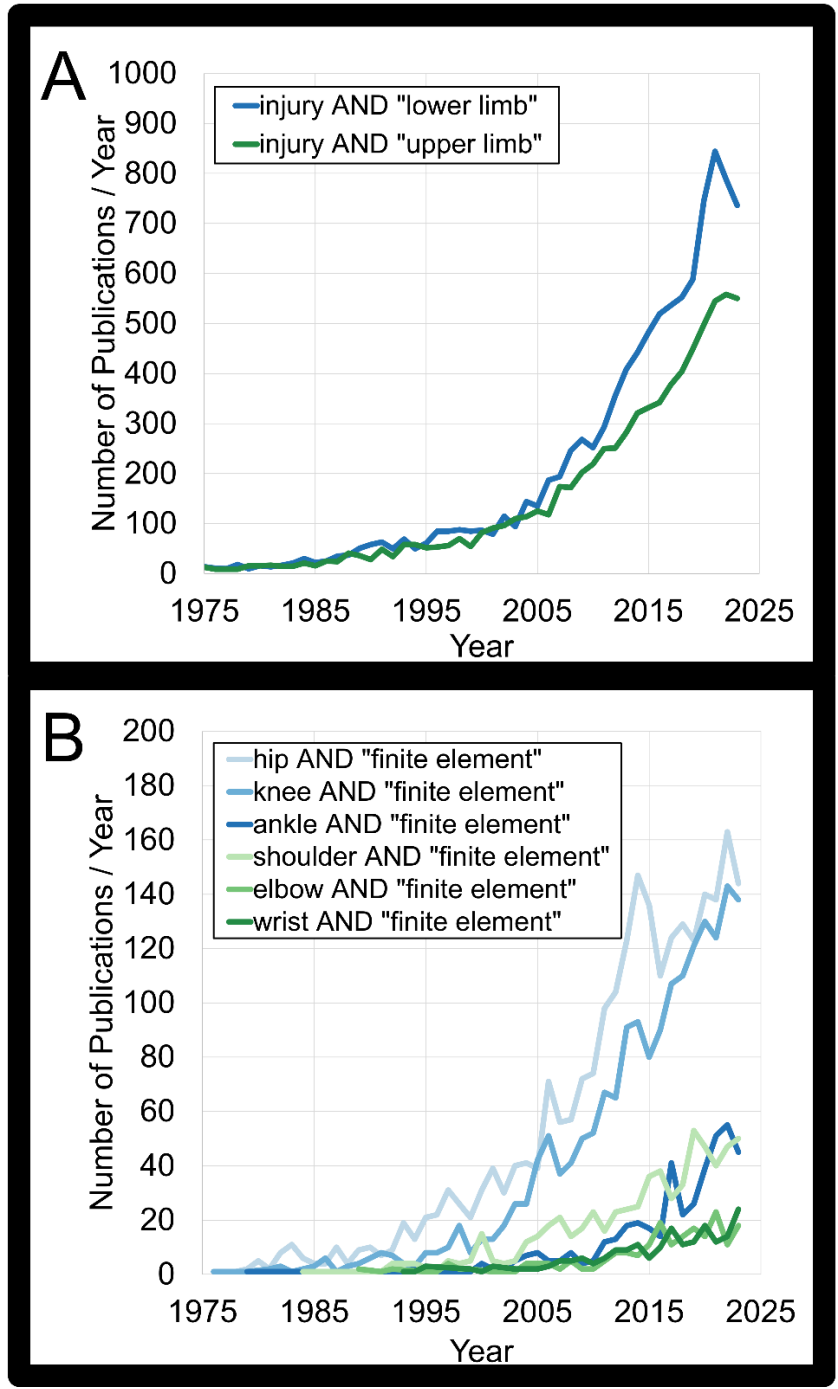

Figure 1: PubMed-indexed articles about lower and upper extremity joints from 1975 to 2024. (A) The number of injury-focused publications is similar between the upper and lower limb. (B) Publications per year focused on "finite element" stratified by joint. The wrist has the least finite element model-based publications of any joint queried.

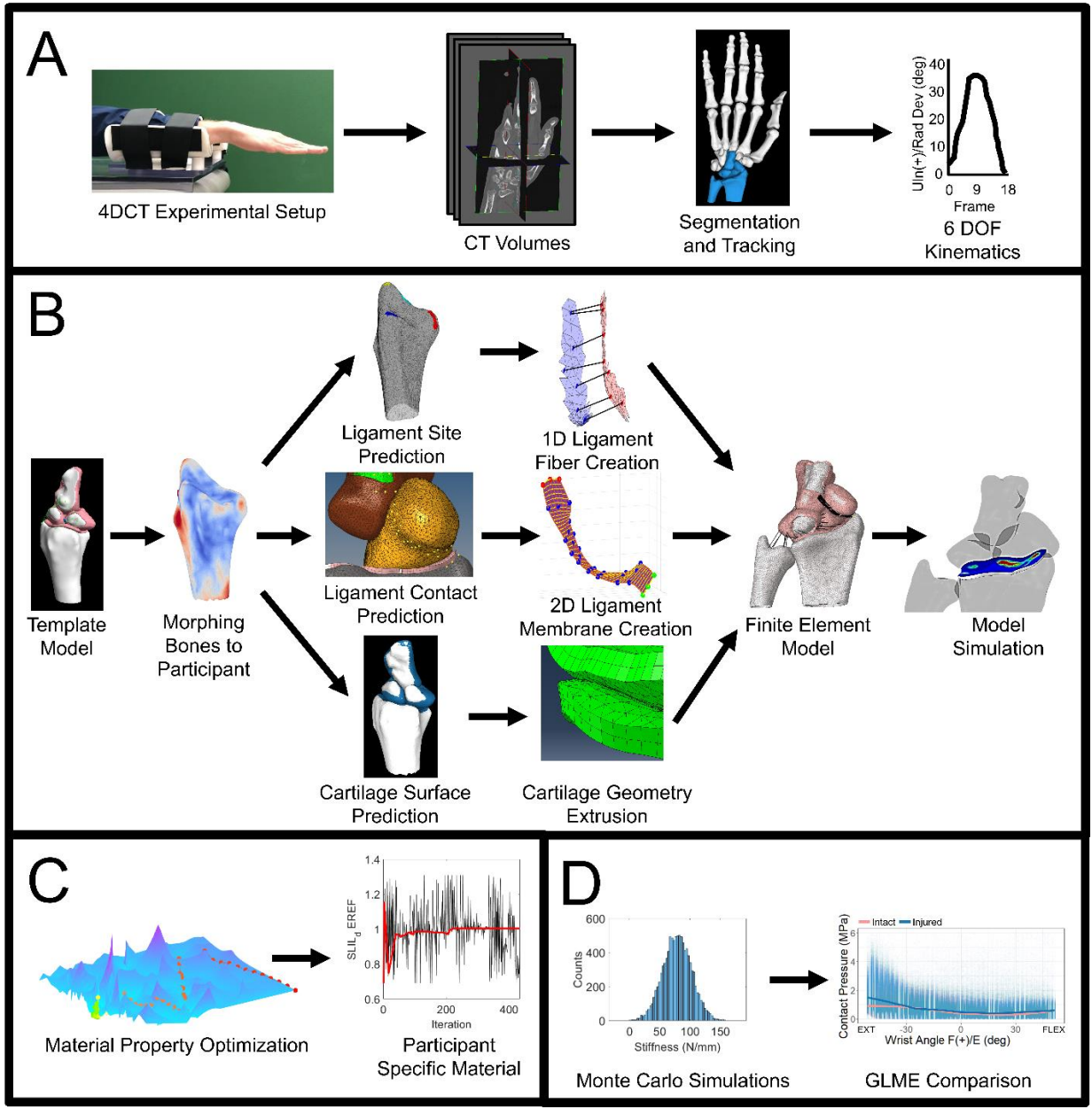

Figure 2: Overall Study Workflow. (A) Experimental workflow from four-dimensional computed tomography (4DCT) to resulting bone surface geometries and 6DOF joint kinematics. (B) Model development workflow using bone surface geometries and a template model to define participant-specific ligament and cartilage geometries to create the final finite element model. (C) Ligament material optimization to obtain participant-specific material properties from minimizing 6 DOF kinematic cost (Analysis 1) (D) Ligament Monte Carlo variation to determine effects of material property variation and ligament injury on joint contact and ligament loads (Analysis 2).

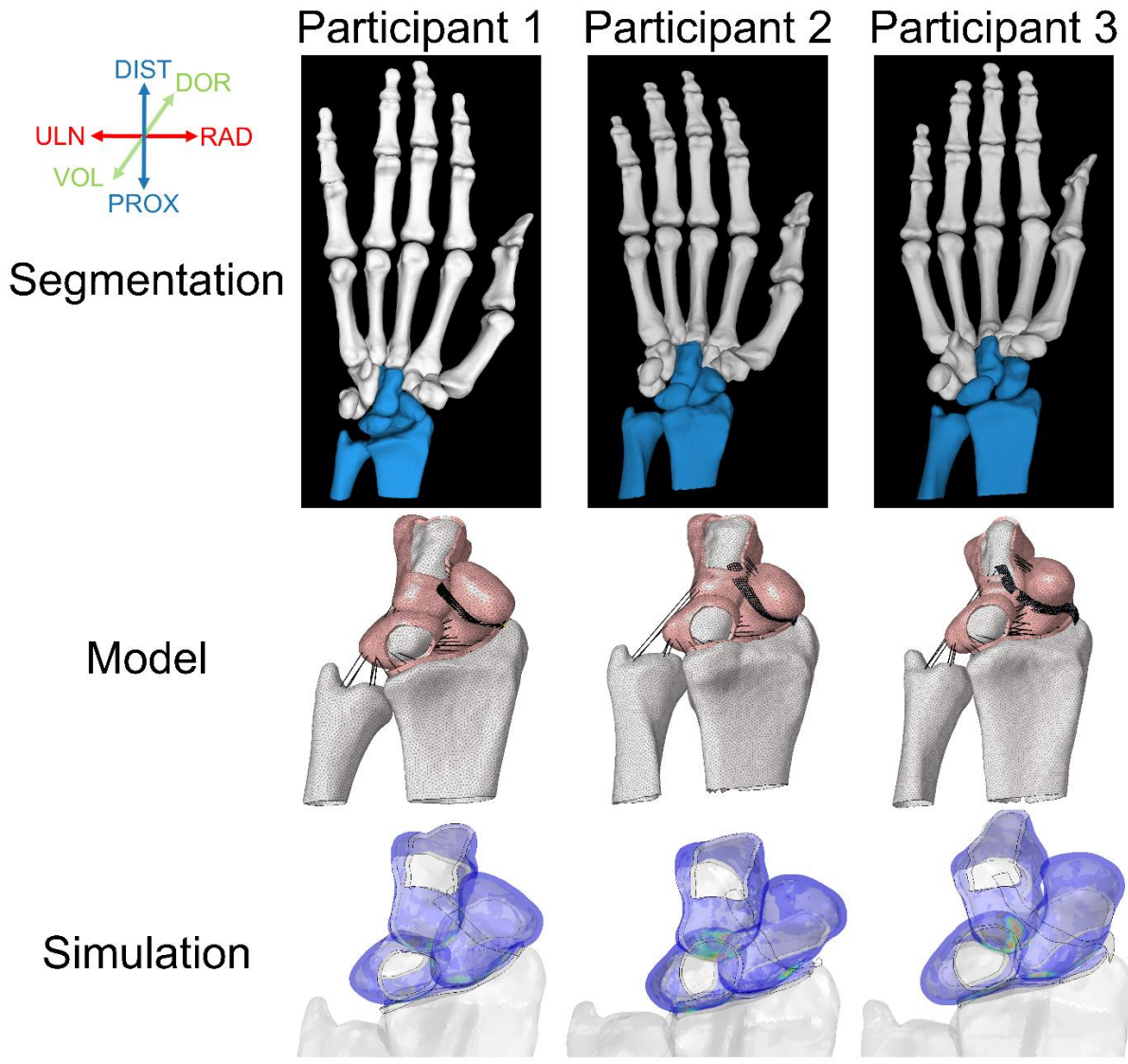

Figure 3: 3D bone surfaces generated from segmentation from static CT, finite element model, and model simulation with a sample frame of the contact pressure plot at the neutral pose, representing a radiocapitate extension angle of approximately 0°.

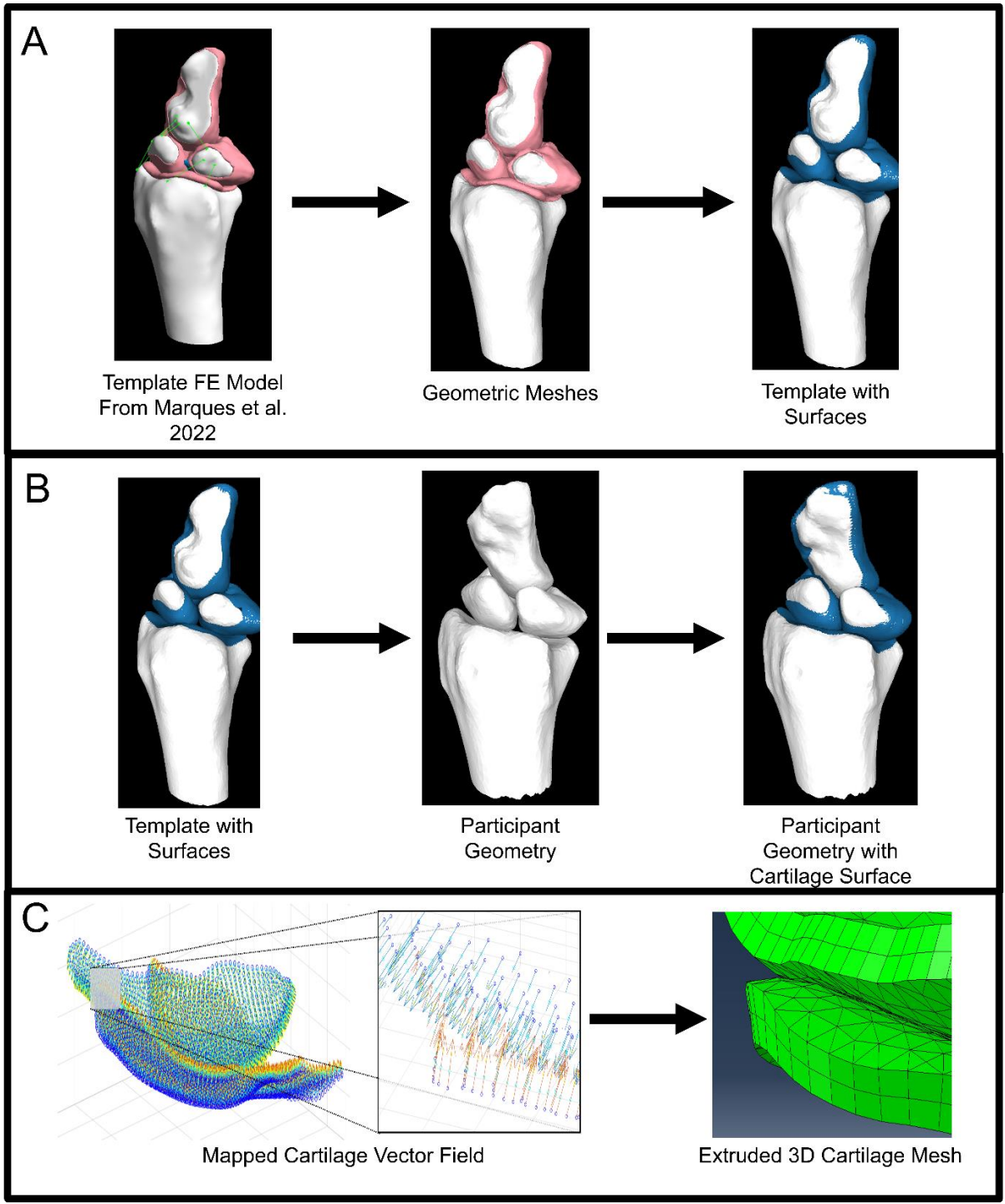

Figure 4: Cartilage 3D mesh creation from approximate surface regions and non-linear mapped extrusion. (A) Creation of cartilage surface (yellow) from projection of original 3D mesh to bone surfaces for template geometry. A template bone from Marques et al., with previously identified cartilage, was used. 3D cartilage geometries were projected based on their surface normal directions to determine the corresponding cartilage-covered bone surface. (B) Creation of approximate cartilage surface on participant geometry via morphing of template to participant bones and then approximating cartilage

region (blue). Template bones with identified cartilage regions were used. These were morphed to each participant's geometry using validated non-linear morphing algorithms; the solution was used to predict the location of the cartilage surfaces on the new participant's bones, based on mapped position from the original template mode. (C) Cartilage thickness extrusion to 3D wedge elements via mapped nearest-neighbor distance vector field. The resulting cartilage surfaces on each pair of bones were used to create a distance field to the nearest cartilaginous surface based on all other bones. The distance was used to create 3D mapped neural network solutions for the cartilage thickness and extrude nodal vertices to create 3D wedge elements.

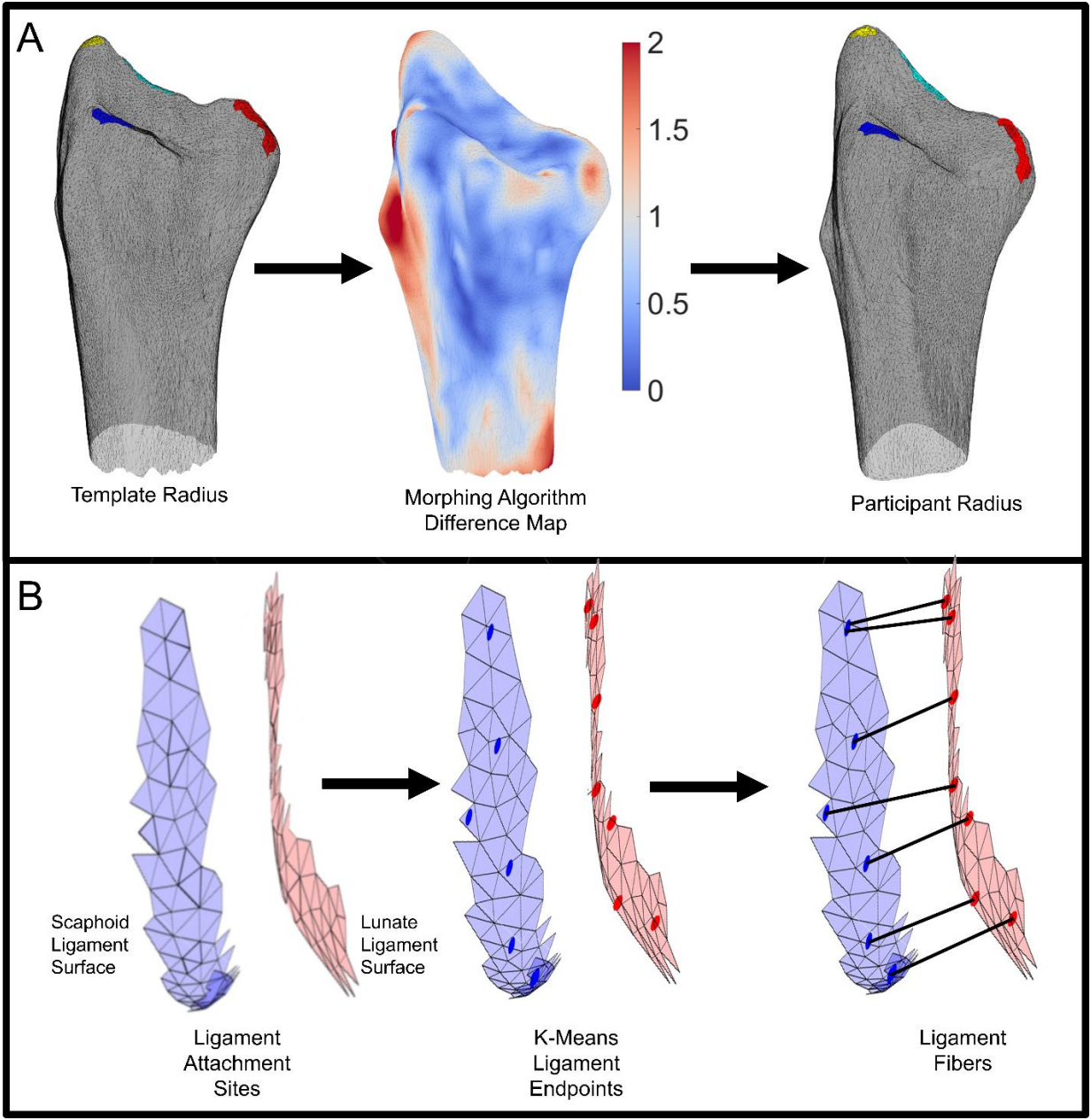

Figure 5: Ligament 1D fiber creation from non-linear morphing and fiber creation. (A) Identification of ligament attachment sites from a template bone followed by non-linear morphing to participant bone. The difference map shows regions of similarity (blue) and dissimilarity (red) between the template and participant bone. Colored regions represent predicted ligament attachment sites. (B) Creation of ligament fibers from ligament attachment sites on two paired bones. Endpoints are found on each surface using K-means methods. Fibers are created using the Kuhn-Munkres method.

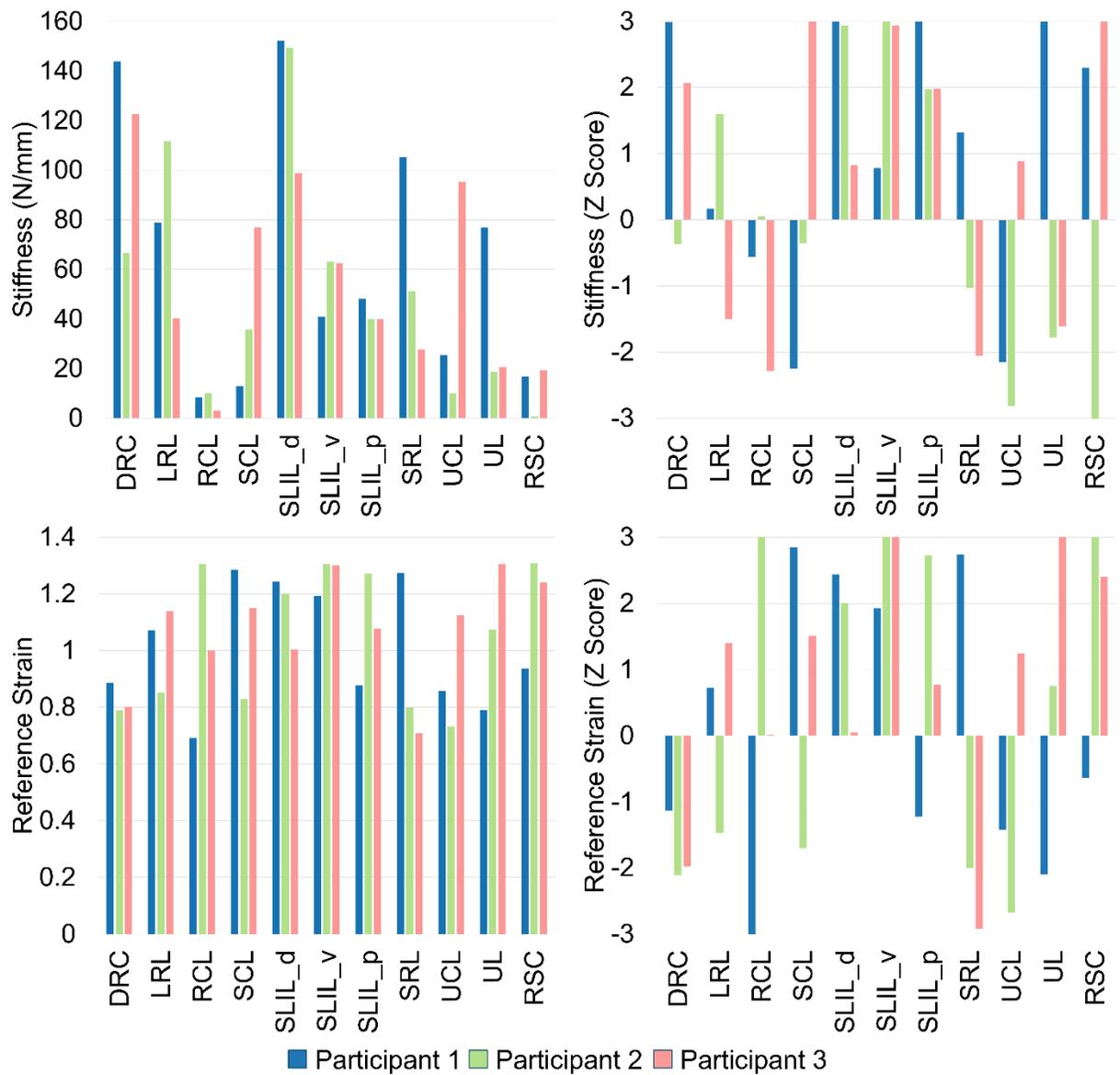

Figure 6: Model-derived optimized ligament material properties. Reference strain values greater than 1.0 represent initial tension in the static CT-derived neutral wrist position. Values reported as true metrics (left) and Z-score normalized to the mean and standard deviation for each material property (right).

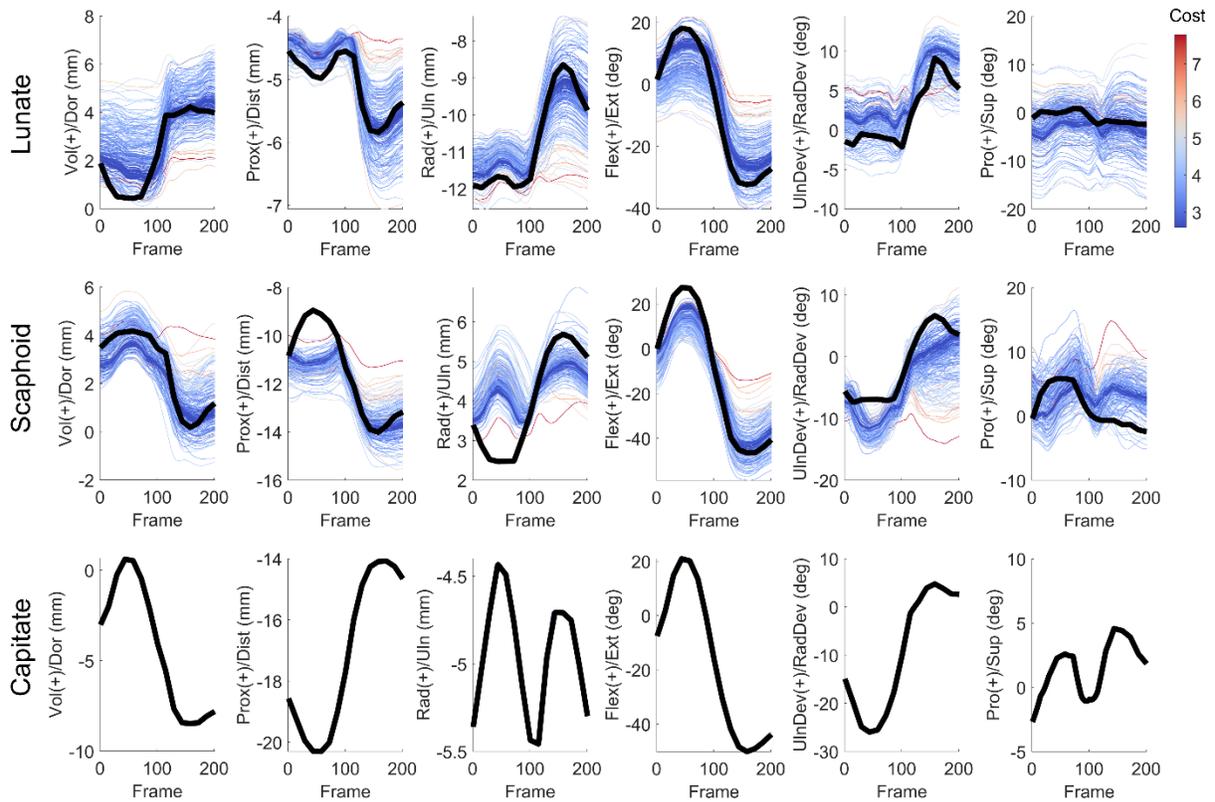

Figure 7: Kinematic predictions from Participant 3 during model optimization of participant unresisted flexion-extension motion. Kinematics were driven for the capitate in displacement control, while the lunate and scaphoid were free to move. Solid black lines represent experimentally observed 4DCT motion. All other lines represent a single optimization iteration, with lines in red having the greatest cost, and lines in blue having the smallest cost. Final optimized ligament parameters for model optimization are from the simulation with the lowest total cost.

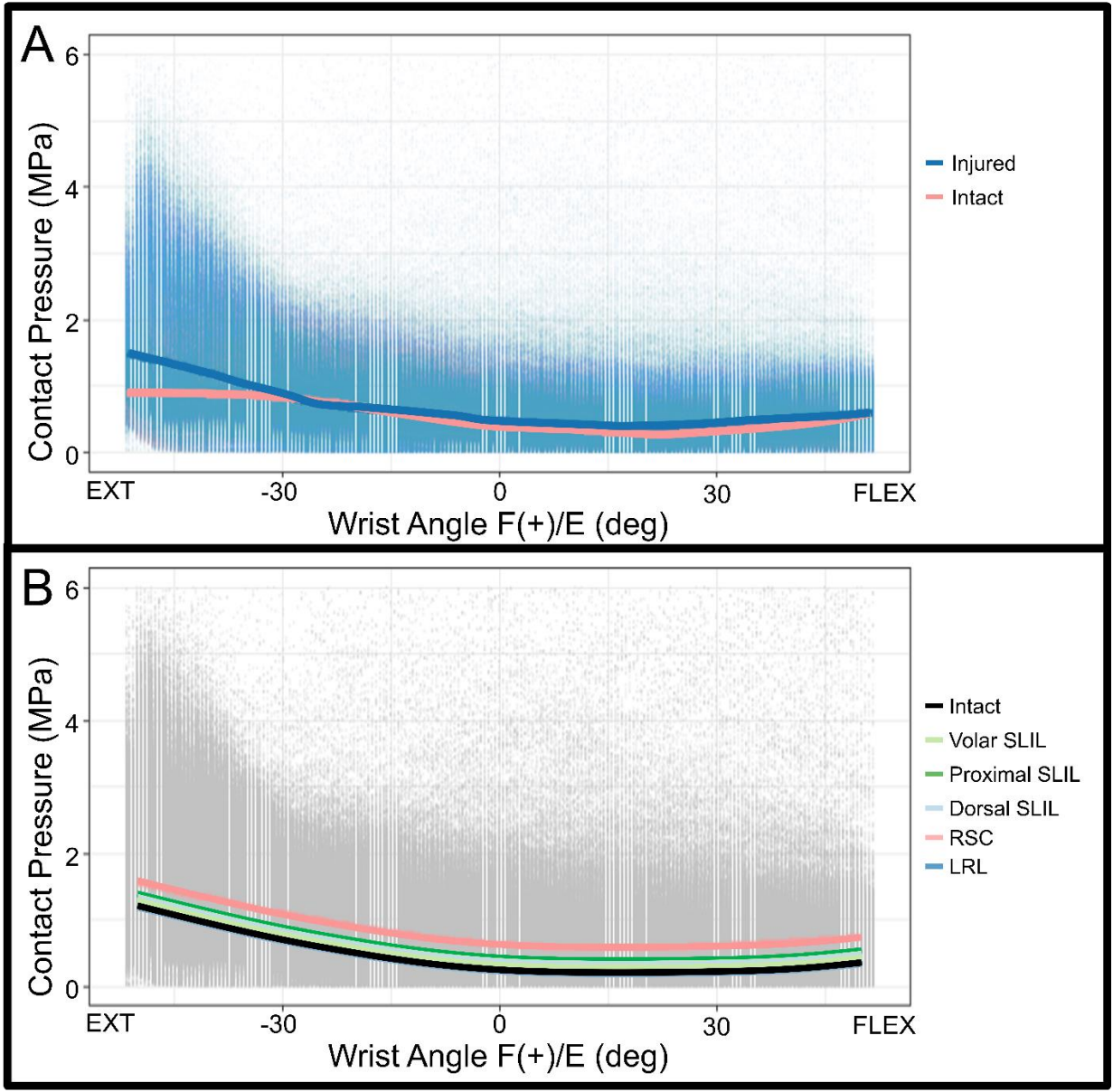

Figure 8: MC simulations predicting average radioscaphoid contact pressure as a function of wrist flexion-extension. (A) Solid lines are medians of pressures at each wrist angle for models without ligament injury (pink) and those with at least one ligament injured (blue). Simulations are shown as individual dots. (B) GLME-estimated average radioscaphoid contact pressure following injury to the labeled ligament. In both graphs, individual simulations are shown as individual dots.

Andreassen et al. Page **31** of **33**

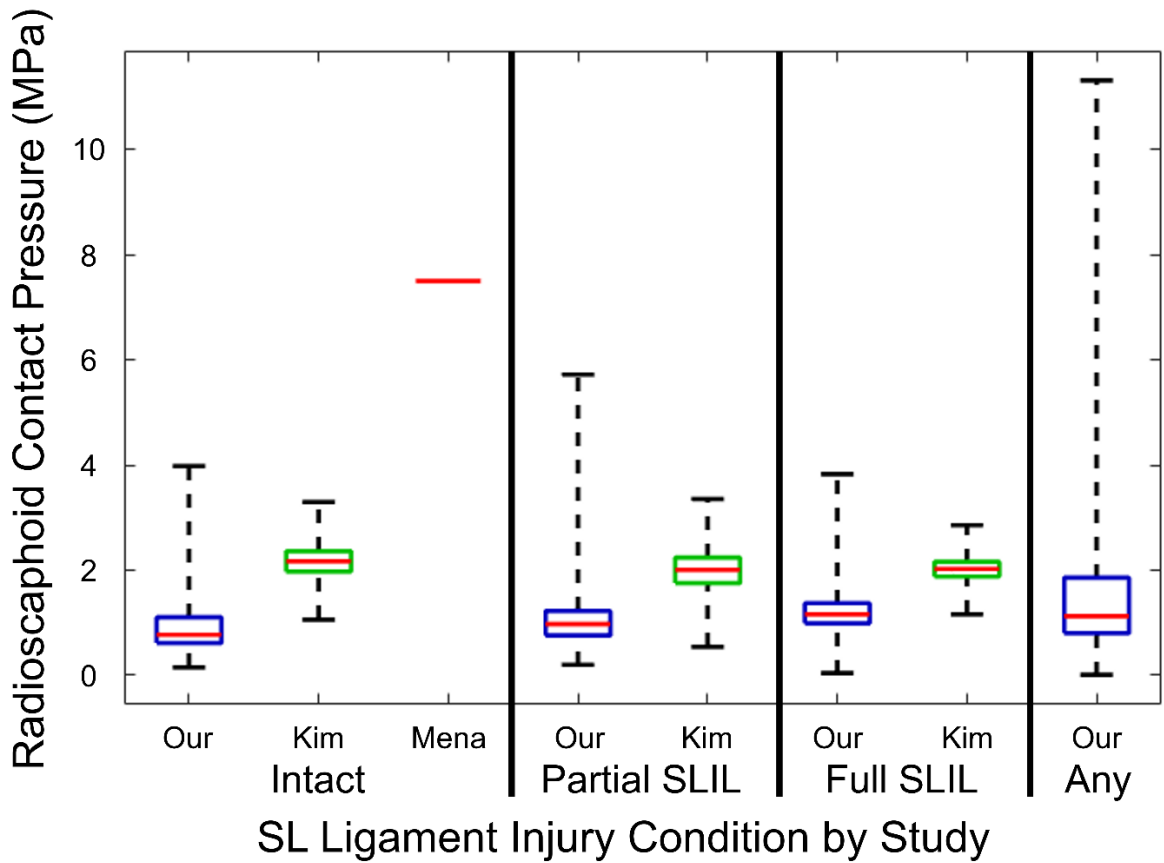

Figure 9: Distribution of radioscaphoid contact pressure from MC analysis at full extension for Participant 1 and Participant 2 compared to values reported in Kim et al. and Mena et al. [11, 14]. "Intact" refers to simulations with all ligaments intact. "Partial SLIL" refers to simulations with at least one bundle of the SLIL injured with no extrinsic ligament injuries. "Full SLIL" refers to simulations wherein all bundles of the SLIL are injured with no extrinsic ligament injury. "Any injury" refers to models with at least one ligament injured (any combination of injury to the volar SLIL, proximal SLIL, dorsal SLIL, RSC, and LRL).

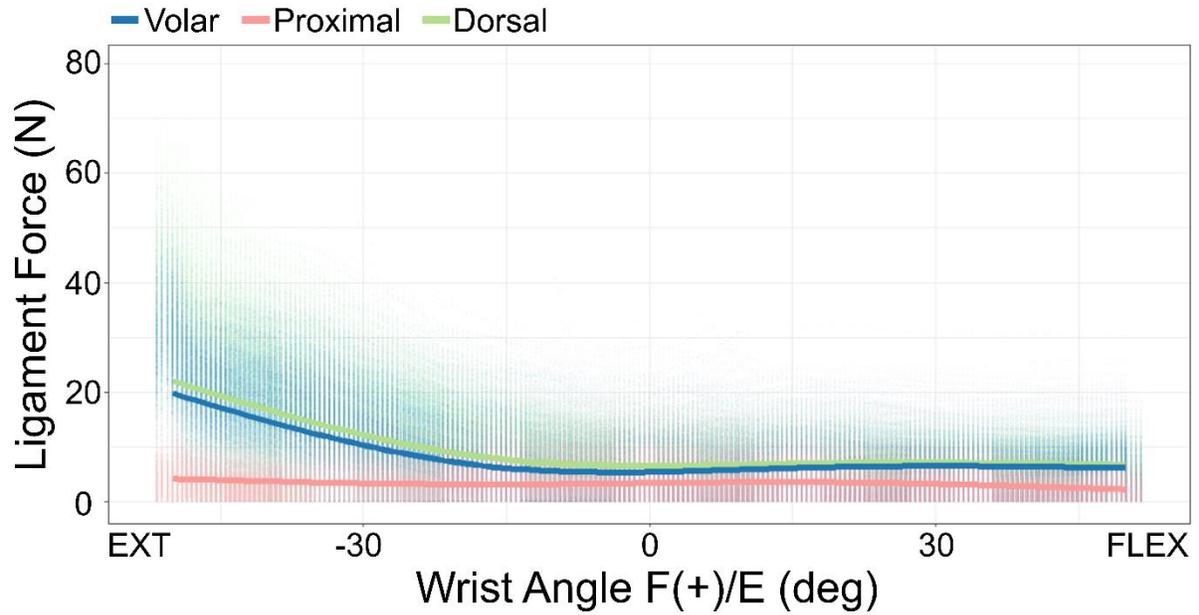

Figure 10: Simulated ligament forces during MC Analysis in SLIL bundles during wrist flexion and extension for Participants 1 and 2 across all intact models. Solid lines are median ligament forces in SLIL bundles across all intact MC simulations. Points are individual MC simulations with all ligaments intact.